\documentstyle[budapest1]{article}
\frompage{000} \topage{000}                                              
\def\rightmark{(Short) The $\pi\, N \rightarrow e^+e^- N$ reaction}
\def\leftmark{M. Soyeur et al.}

\title{The $\pi\, N \rightarrow e^+e^- N$ reaction \\
\smallskip
close to the vector meson production threshold $^{a}$}
\authors{
{Madeleine Soyeur$^1$, Matthias F.M. Lutz$^2$ and Bengt Friman$^{2}$%
}\\[2.812mm]
{\normalsize
\hspace*{-8pt}$^1$ DAPNIA/SPhN,
CEA/Saclay, F-91191 Gif-sur-Yvette Cedex, France\\[0.2ex]
\hspace*{-8pt}$^2$ GSI, Planckstrasse 1,  D-64291 Darmstadt, Germany\\
Institut f\"ur Kernphysik, TU Darmstadt,
   D-64289 Darmstadt, Germany
}}

\abstract{The $\pi^-p \rightarrow e^+e^- n$ and $\pi^+n \rightarrow e^+e^- p$
reaction cross sections are calculated below and in the vicinity
of the vector meson ($\rho^0$, $\omega$) production threshold.
These processes are largely responsible for the emission of
$e^+e^-$ pairs in pion-nucleus reactions and contribute to the
dilepton spectra observed in relativistic heavy ion collisions.
They are dominated by the decay of low-lying baryon resonances
into vector meson-nucleon channels. The vector mesons materialize subsequently
into $e^+e^-$ pairs. Using $\pi N\rightarrow \rho^0 N$ and
$\pi N \rightarrow \omega N$ amplitudes calculated in the
center of mass energy interval $1.4<\sqrt s <1.8$ GeV, we compute
the $\pi^-p \rightarrow e^+e^- n$ and $\pi^+n \rightarrow e^+e^- p$
reaction cross sections in these kinematics. Below the vector meson production threshold, the
$\rho^0 - \omega$ interference in the $e^+e^-$ channel
appears largely destructive for
the $\pi^-p \rightarrow e^+e^- n$ cross section and
constructive for the $\pi^+n \rightarrow e^+e^- p$ cross section.
The pion beam and the HADES detector
at GSI offer a unique possibility to measure these effects. Such data would provide
strong constraints on the coupling of vector meson-nucleon channels to low-lying baryon
resonances.}

\begin{document}

\maketitle

\setcounter{page}{1}

\section{Introduction}\label{intro}

The study of the $\pi^-p \rightarrow e^+e^- n$ and $\pi^+ n \rightarrow e^+e^- p$ processes described
in this work \cite{Lutz2} aims at gaining understanding of the
$\pi N \rightarrow \rho^0 N$ and $\pi N \rightarrow \omega N$
scattering amplitudes for center of mass ener\-gies close and below the
vector meson production threshold ($1.5<\sqrt s <1.8$ GeV) \cite{Lutz1}.
There are well-known baryon resonances in
this energy range, which contribute
to the $\pi^-p \rightarrow e^+e^- n$ and $\pi^+ n \rightarrow e^+e^- p$ scattering

\medskip
\smallskip
\hrule width 15mm
\begin{notes}
\item[a]
This talk is based on the work published in Ref. [1].
\end{notes}
amplitudes through their coupling to the $\pi N$, $\rho^0 N$ and $\omega N$ channels.
These amplitudes involve in addition significant non-resonant processes.
The phenomenological $\rho N N^*$ and $\omega N N^*$
coupling strengths needed to understand the data related to the
$\pi N \rightarrow \rho^0 N$ and $\pi N \rightarrow \omega N$
amplitudes are pivotal quantities
for baryon structure studies \cite{Lutz1}.

The exclusive observation of neutral vector mesons through their $e^+e^-$
decay presents definite advantages over their observation
through final states invol\-ving pions.
Firstly, there are no competing processes, such as $\pi \Delta$ production which leads to the
same $\pi \pi N$ final state and
impairs consequently the identification of the $\rho$-meson in that channel. Se\-condly, both the $\rho^0$- and
$\omega$-mesons decay into the $e^+e^-$ channel.
This leads to a quantum interference pattern which is expected to reflect sensitively
the structure and relative sign of the $\pi N \rightarrow \rho^0 N$
and $\pi N \rightarrow \omega N$ scattering amplitudes.

A proper understanding of the
$\pi^-p \rightarrow e^+e^- n$ and $\pi^+ n \rightarrow e^+e^- p$
reactions appears also as a necessary step
towards a detailed interpretation of the production of lepton pairs off nuclei
induced by charged pions. Such reactions would be particularly sensitive to
the propagation of $\omega$-mesons in nuclei \cite{Schoen}.

In Section 2, we present the relativistic coupled-channel model \cite{Lutz1}
used to des\-cribe the $\pi N \rightarrow \rho^0 N$ and $\pi N
\rightarrow \omega N$ amplitudes and outline the calculation of the
$\pi^-p \rightarrow e^+e^- n$ and
$\pi^+n \rightarrow e^+e^- p$
cross sections in the
Vector Meson Dominance model. Our numerical results
for these cross sections
are displayed in Section 3.
We discuss the $\rho^0-\omega$ quantum interference pattern in the
$e^+e^-$ spectrum for both the $\pi^-p \rightarrow e^+e^- n$ and $\pi^+ n \rightarrow e^+e^- p$
reactions. We conclude briefly in Section 4.

\section{Calculation of the $\pi^-p \rightarrow e^+e^- n$ and $\pi^+n \rightarrow e^+e^- p$ cross sections
close to the vector
meson production threshold}

We describe the $\pi N \rightarrow e^+e^- N$ reaction for
$e^+e^-$ pair invariant masses ran\-ging from $\sim$0.4 to $\sim$0.8 GeV.
Assuming Vector Meson Dominance for the electromagnetic current \cite{Kroll},  the
$\pi N \rightarrow \rho^0 N$ and $\pi N
\rightarrow \omega N$ amplitudes are the basic
quantities entering the calculation of the $\pi N \rightarrow e^+e^- N$
cross section. This assumption is illustrated in Fig. 1, where we
show the diagrams contributing to the $\pi^-p \rightarrow e^+e^- n$
process.
\begin{figure}[t]
                 \insertplot{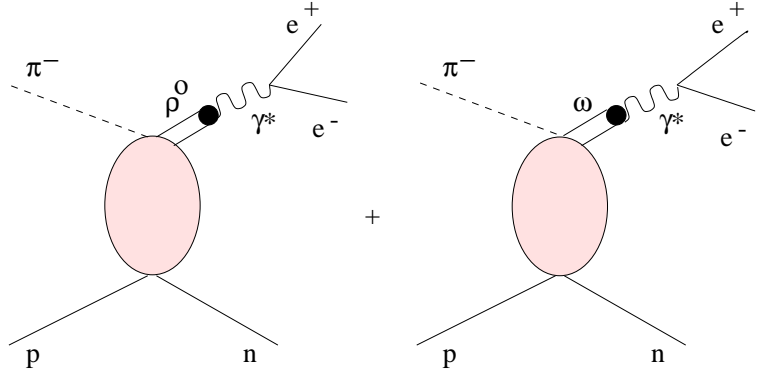}
\caption[]{Diagrams contributing to the $\pi^-p \rightarrow  e^+e^- n$
amplitude with intermediate $\rho^0$- and $\omega$-mesons.}
\label{fig1}
\end{figure}

We use the $\pi N \rightarrow \rho^0 N$ and $\pi N \rightarrow \omega N$
amplitudes obtained in the recent relativistic and unitary coupled-channel
approach to
meson-nucleon scattering of Ref. \cite{Lutz1}.
The available data on pion-nucleon elastic and inelastic scatte\-ring and
on meson photoproduction off nucleon targets
are fitted in the energy window $1.4<\sqrt s <1.8$ GeV, using an effective
Lagrangian with quasi-local two-body meson-baryon interactions
and a generalized form of Vector Meson Dominance to describe the coupling
of vector mesons to real photons.
The scheme comprises the $\pi N$, $\pi \Delta$, $\rho N$,
$\omega N$, $K \Lambda$, $K \Sigma$ and $\eta N$ hadronic channels.
The coupling cons\-tants entering the effective Lagrangian are parameters
which are adjusted to reproduce the data. In view of the kinematics,
only s-wave scattering in the $\rho N$ and $\omega N$ channels is included,
restricting $\pi N$ and $\pi \Delta$ scattering to s- and d-waves.
The pion-nucleon resonances in the S$_{11}$, S$_{31}$, D$_{13}$ and D$_{33}$
partial waves are generated dynamically by solving Bethe-Salpeter
equations \cite{Lutz1}. In the $\rho^0 N$- and $\omega N$-channels,
the restriction to s-wave scattering means that the model applies
to situations where the vector meson is basically at rest with
respect to the scattered nucleon $(\sqrt s \simeq M_N + M_V)$.
This assumption implies that the range of validity of the present
calculation is limited to $e^+e^-$ pairs with invariant masses
$m_{e^+e^-}$ close to $(\sqrt s - M_N)$ and to values of $\sqrt s$
below and
very close to threshold i.e. $1.5 <\sqrt s \leq 1.75$ GeV.

The $\pi N \rightarrow \rho N$ amplitude has isospin 1/2
and isospin 3/2 components while the  $\pi N \rightarrow \omega N$ amplitude
selects the isospin 1/2 channel. Both amplitudes have spin 1/2 and spin 3/2
parts.

The invariant transition matrix elements for the $\pi N \rightarrow \rho N$
and $\pi N \rightarrow \omega N$ reactions are given by

\begin{eqnarray}
\langle \rho^j (\overline{q})\,N(\overline{p})|\,{\mathcal T}\,| \pi^i(q)\, N(p)\rangle & \nonumber \\
= \,(2\pi)^4\,  \delta^4 (q+ &p- \overline{q}- \overline{p})\, \overline u(\overline{p})\, \epsilon^\mu (\overline{q})
\,T^{ij}_{(\pi N\rightarrow \rho N)\,\mu}\,u(p),
\label{eq:e1}
\end{eqnarray}
\begin{eqnarray}
\langle \omega (\overline{q})\,N(\overline{p})|\,{\mathcal T}\,| \pi^i(q)\, N(p)\rangle & \nonumber \\
= \,(2\pi)^4\,  \delta^4 (q+ &p- \overline{q}- \overline{p})\, \overline u(\overline{p})\, \epsilon^\mu (\overline{q})
\,T^i_{(\pi N\rightarrow \omega N)\,\mu}\,u(p),
\label{eq:e2}
\end{eqnarray}

\noindent
where $T^{ij}_{(\pi N\rightarrow \rho N)\,\mu}$ and
$T^{i}_{(\pi N\rightarrow \omega N)\,\mu}$ are functions of the three kinematic variables
$w=p+q=\overline{p}+\overline{q}$ ($\sqrt {w^2}$ = $\sqrt s$), $q$ and
$\overline{q}$.
These scattering amplitudes can be decomposed into isospin invariant components
as
\begin{eqnarray}
T^{ij}_{(\pi N\rightarrow \rho N)\, \mu} (\overline{q},q;w)
\, = \sum_{I}
T^{(I)}_{(\pi N \rightarrow \rho N)\,\mu}(\overline{q},q;w)\, P_{(\rho)}^{(I)\,ij},
\label{eq:e5}
\end{eqnarray}
\begin{eqnarray}
T^{i}_{(\pi N\rightarrow \omega N)\, \mu} (\overline{q},q;w)
\, = \sum_{I}
T^{(I)}_{(\pi N \rightarrow \omega N)\, \mu}\,(\overline{q},q;w) P_{(\omega)}^{(I)\,i},
\label{eq:e6}
\end{eqnarray}
\noindent
in which $P_{(\rho)}^{(I)\,ij}$ and $P_{(\omega)}^{(I)\,i}$ are isospin projectors \cite{Lutz2}.
The isospin invariant amplitudes can be expanded further into
components of total angular momentum,

\begin{eqnarray}
T^{(I)}_{(\pi N\rightarrow V N)\, \mu} (\overline{q},q;w)
\,& = \,
M^{(I,J=\frac {1} {2})}_{\pi N \rightarrow V N}(s)\,
\,{Y}_{(J=\frac {1} {2})\,\mu}(\overline{q},q;w)\nonumber \\
&+ \, M^{(I,J=\frac {3} {2})}_{\pi N \rightarrow V N}(s)\,
\,{Y}_{(J=\frac {3} {2})\,\mu}(\overline{q},q;w).
\label{eq:e10}
\end{eqnarray}
\noindent
V stands for $\rho$ or $\omega$ and ${Y}_{(J=\frac {1} {2})\,\mu}(\overline{q},q;w)$
and ${Y}_{(J=\frac {3} {2})\,\mu}(\overline{q},q;w)$
are relativistic angular momentum
projectors \cite{Lutz2}.
\par
The $\pi N \rightarrow \rho N$ and $\pi N \rightarrow
\omega N$ amplitudes in the S$_{11}$, S$_{31}$, D$_{13}$ and D$_{33}$
channels obtained in Ref. \cite{Lutz1} are  displayed
in Figs. 2 and 3. The quantities shown
are the amplitudes $M^{(I,J)}_{\pi N\rightarrow \rho N}(s)$
and $M^{(I,J)}_{\pi N\rightarrow \omega N}(s)$ defined by
Eq. (\ref {eq:e10}), which depend only on the center of mass
energy $\sqrt s$.

\begin{figure}[h]
\vspace*{-0.5 cm}
                 \insertplot{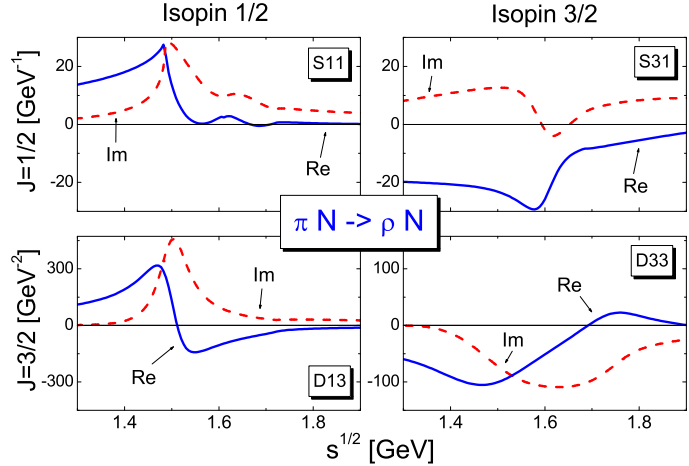}
\vspace*{-0.9 cm}
\caption[]{Real and imaginary parts of
the $\pi N \rightarrow \rho^0 N$ amplitudes
in the pion-nucleon
S$_{11}$, S$_{31}$, D$_{13}$ and D$_{33}$
partial waves [1].}
\label{fig2}
\end{figure}
The coupling to subthreshold resonances is clearly
exhi\-bited in these pictures.
In the S$_{11}$ channel, the N(1535) and the N(1650) resonances lead
to peak structures in the imaginary parts of the amplitudes.
The pion-induced $\omega$ production amplitudes in the D$_{13}$ channel
reflect the strong coupling of the N(1520) re\-sonance to the
$\omega$N channel.
\begin{figure}[h]
\vspace*{-0.5 cm}
                 \insertplot{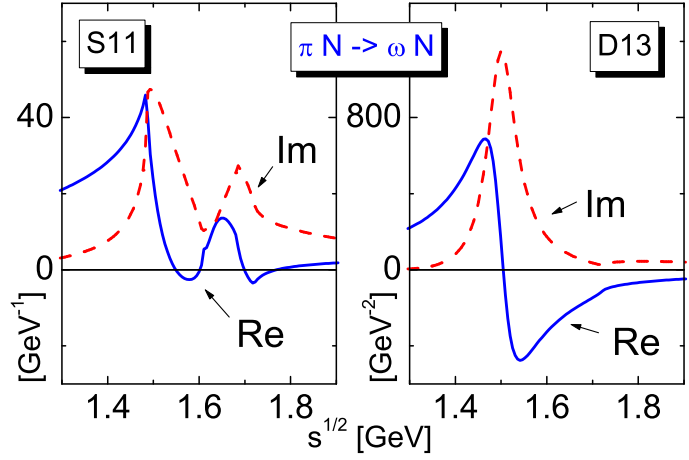}
\vspace*{-0.9 cm}
\caption[]{Real and imaginary parts of
the $\pi N \rightarrow \omega N$ amplitudes in the pion-nucleon
S$_{11}$ and D$_{13}$ partial waves [1].}
\label{fig3}
\end{figure}
The $\pi^-p \rightarrow \rho^0 n$ and $\pi^-p \rightarrow \omega n$
amplitudes are obtained from the
isospin 1/2 and isospin 3/2 scattering amplitudes by the
relations,
\noindent
\begin{eqnarray}
M^J_{\pi^-p \rightarrow \rho^0 n} = - \frac {\sqrt 2} {3}
M^{(1/2,J)}_{\pi N\rightarrow \rho N}  +\frac {\sqrt 2} {3}
M^{(3/2,J)}_{\pi N\rightarrow \rho N},
\label{eq:e17}
\end{eqnarray}
\noindent
\begin{eqnarray}
M^J_{\pi^-p \rightarrow \omega n} = \sqrt{\frac {2} {3}}
M^{(1/2,J)}_{\pi N\rightarrow \omega N}.
\label{eq:e18}
\end{eqnarray}

\noindent
Similarly the $\pi^+ n \rightarrow \rho^0 p$ and $\pi^+n \rightarrow \omega p$
amplitudes are given by
\noindent
\begin{eqnarray}
M^J_{\pi^+n \rightarrow \rho^0 p} = \frac {\sqrt 2} {3}
M^{(1/2,J)}_{\pi N\rightarrow \rho N}  -\frac {\sqrt 2} {3}
M^{(3/2,J)}_{\pi N\rightarrow \rho N},
\label{eq:e19}
\end{eqnarray}
\noindent
\begin{eqnarray}
M^J_{\pi^+n \rightarrow \omega p} = \sqrt{\frac {2} {3}}
M^{(1/2,J)}_{\pi N\rightarrow \omega N}.
\label{eq:e20}
\end{eqnarray}

\newpage
The phases of the isospin coefficients appearing in Eqs. (\ref {eq:e17}) and (\ref {eq:e19}) play a crucial
role in determining the $\rho^0-\omega$ interference in the
$\pi^-p \rightarrow e^+e^- n$ and $\pi^+n \rightarrow e^+e^- p$
reaction cross sections.
The real and imaginary parts of the $\pi^-p \rightarrow \omega n$
and of the $\pi^+n \rightarrow \omega p$ amplitudes are the same and mostly positive.
In contrast, the $\pi^-p \rightarrow \rho^0 n$ and $\pi^+n \rightarrow \rho^0 p$ amplitudes
have opposite signs. The $\pi^-p \rightarrow \rho^0 n$ amplitudes are predominantly
negative and will therefore interfere destructively with the $\pi^-p \rightarrow \omega n$
amplitudes. The $\pi^+n \rightarrow  \rho^0 p$ and $\pi^+n \rightarrow \omega p$
amplitudes have the same sign over a large $\sqrt s$ interval, leading to a constructive
interference.

The $\pi^-p \rightarrow e^+e^- n$ and $\pi^+n \rightarrow e^+e^- p$ cross sections are calculated from
the $\pi^-p \rightarrow \rho^0 n$, $\pi^-p \rightarrow \omega n$,
$\pi^+n \rightarrow \rho^0 p$ and $\pi^+n \rightarrow \omega p$
amplitudes, assuming Vector Meson Dominance of the electromagnetic current \cite{Sakurai,Kroll}.
This assumption can be enforced in the effective Lagrangian by introducing
vector meson-photon interaction terms of the form,
\begin{eqnarray}
{\mathcal L^{int}_{\gamma V}}\,&=&\, \frac {f_\rho} {2 M_\rho^2} F^{\mu \nu}\, \rho^0_{\mu \nu}
\,+\,\frac {f_\omega} {2 M_\omega^2} F^{\mu \nu}\, \omega_{\mu \nu},
\label{eq:e21}
\end{eqnarray}
\noindent
where the photon and vector meson field tensors are defined by
\begin{eqnarray}
F^{\mu \nu}\, =\,\partial^\mu A^\nu -\partial^\nu A^\mu,
\label{eq:e22}
\end{eqnarray}
\begin{eqnarray}
V^{\mu \nu}\, =\,\partial^\mu V^\nu -\partial^\nu V^\mu.
\label{eq:e23}
\end{eqnarray}
In equation (\ref {eq:e21}),
$M_\rho$ and $M_\omega$ are the $\rho$- and $\omega$-masses and
$f_\rho$ and  $f_\omega$ are dimensional coupling constants. Their magnitude can be
determined from the $e^+e^-$ partial decay widths of the $\rho$- and
$\omega$-mesons to be \cite{Friman1}
\begin{eqnarray}
|f_\rho|= 0.036\, GeV^2,
\label{eq:e24}
\end{eqnarray}
\begin{eqnarray}
|f_\omega|= 0.011 \,GeV^2.
\label{eq:e25}
\end{eqnarray}
The relative sign of $f_\rho$ and $f_\omega$
is fixed by vector meson photoproduction amplitudes \cite{Lutz1}.
We assume that the phase correlation between isoscalar and isovector currents
is identical for real and virtual photons as in Sakurai's realization
of the Vector Meson Dominance assumption \cite{Sakurai}. With the conventions used
in this paper, both $f_\rho$ and $f_\omega$ are positive.

\section{Numerical results}

With the $\pi N \rightarrow \rho N$ and $\pi N \rightarrow
\omega N$ amplitudes and the Vector Meson Dominance assumption
discussed in Section 2, we have calculated the differential
cross section $\frac{d\sigma}{d\overline{q}^2}_{\pi N \rightarrow e^+e^- N}$
for the $\pi^-p \rightarrow e^+e^- n$ and $\pi^+n \rightarrow e^+e^- p$
reactions. The magnitude of the 4-vector $\overline{q}$
is the invariant mass m$_{e^+e^-}$ of the $e^+e^-$ pair.
We refer to \cite{Lutz2} for calculational details.

The differential cross sections for the $\pi^-p \rightarrow e^+e^- n$
and the $\pi^+n \rightarrow e^+e^- p$ reactions are
computed for values of the total center of mass energy $\sqrt s$
ranging from 1.5 GeV up to 1.75 GeV. We explore the dependence of the $\rho^0-\omega$ interference
pattern in the $e^+e^-$ channel on $\sqrt s$ in this energy range, in particular
in the vicinity of the $\omega$-meson production threshold ($\sqrt s$=1.72 GeV).
We illustrate our results below threshold by displaying in Figs. 4 and 5
the differential cross sections for the $\pi^-p \rightarrow e^+e^- n$
and the $\pi^+n \rightarrow e^+e^- p$ reactions at $\sqrt s$=1.5 GeV,
where the N(1520) and N(1535) baryon
resonances play a dominant role.
These figures show very clearly
the isospin effects discussed in Section 2.
For the two reactions, the $\omega$ and $\rho^0$ contributions to the cross section are the
same.
The $\rho^0$-$\omega$ interference is destructive for the $\pi^-p \rightarrow e^+e^- n$
reaction and constructive for the $\pi^+n \rightarrow e^+e^- p$ process.
Consequently, the $\pi^-p \rightarrow e^+e^- n$ differential cross section is extremely small
in the range of invariant masses considered in this calculation.
In contrast, the constructive $\rho^0$-$\omega$ interference for the $\pi^+n \rightarrow e^+e^- p$
reaction leads to a sizeable differential cross section.
This is a very striking prediction, linked to the resonant structure of the scattering
amplitudes $M^{1/2}_{\pi N\rightarrow V N}$ and
$M^{3/2}_{\pi N\rightarrow V N}$.
Data on differential cross sections for the $\pi^-p \rightarrow e^+e^- n$
and $\pi^+n \rightarrow e^+e^- p$ reactions at $\sqrt s$=1.5 GeV would be very
useful for making progress in the understanding
of the couplings of both the N(1520) and N(1535) baryon
resonances to the vector meson-nucleon channels.
\begin{figure}[h]
\vspace*{-0.5 cm}
                 \insertplot{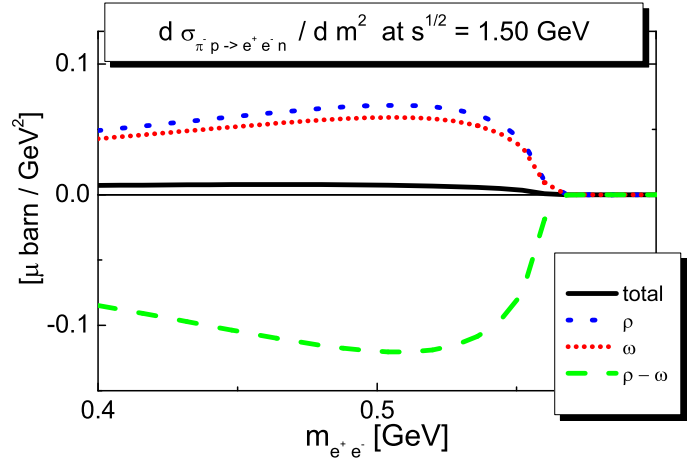}
\vspace*{-0.9 cm}
\caption[]{Differential cross section for the $\pi^-p \rightarrow e^+e^- n$
reaction at $\sqrt s$=1.5 GeV as function of the invariant mass of the $e^+e^-$
pair. The $\rho^0$ and the $\omega$ contributions are indicated by short-dashed
and dotted lines respectively. The long-dashed line shows the $\rho^0-\omega$ interference.
The solid line is the sum of the three contributions.}
\label{fig4}
\end{figure}
\begin{figure}[t]
\vspace*{-0.1 cm}
                 \insertplot{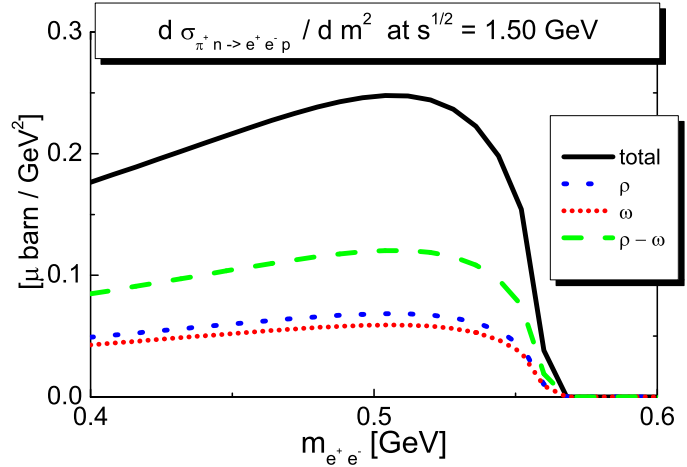}
\vspace*{-1 cm}
\caption[]{Differential cross section for the $\pi^+n \rightarrow e^+e^- p$
reaction at $\sqrt s$=1.5 GeV as function of the invariant mass of the $e^+e^-$
pair. The $\rho^0$ and the $\omega$ contributions are indicated by short-dashed
and dotted lines respectively. The long-dashed line shows the $\rho^0-\omega$ interference.
The solid line is the sum of the three contributions.}
\label{fig5}
\end{figure}

\newpage
The differential
cross sections for the $\pi^-p \rightarrow e^+e^- n$ and $\pi^+n \rightarrow e^+e^- p$
reactions below threshold have been calculated also at $\sqrt s$=1.55, 1.60, 1.65 and 1.70 GeV \cite{Lutz2}.
The cross sections vary smoothly with the total center of mass energy. They
exhibit the features discussed for $\sqrt s$=1.5 GeV,  reflecting however
dynamics associated with higher-lying resonances.
Just below threshold ($\sqrt s$=1.70 GeV), the $\omega$-contribution begins to increase, while the general
features of the $e^+e^-$ production in the two isospin channels remain the same.

The interference pattern changes drastically above the $\omega$-meson threshold.
 Figs. 6 and 7 show the $\pi^-p \rightarrow e^+e^- n$
and $\pi^+n \rightarrow e^+e^- p$ differential cross sections at $\sqrt s$=1.75 GeV.
At this energy,
the differential cross sections for the $\pi^-p \rightarrow e^+e^- n$ and
$\pi^+n\rightarrow e^+e^- p$ reactions are completely dominated by the
$\omega$-contribution. The magnitudes of the cross sections for the two reactions are
comparable. The $\rho^0-\omega$ interference is still destructive
in the $\pi^-p \rightarrow e^+e^- n$ channel and constructive in
the $\pi^+n\rightarrow e^+e^- p$ channel, albeit very small.
In both reactions, crossing the $\omega$-production threshold
leads to a sharp increase in the cross section, by two orders of magnitude
in the $\pi^-p \rightarrow e^+e^- n$ channel and by one order of
magnitude in the $\pi^+n\rightarrow e^+e^- p$ channel.

\newpage

\begin{figure}[h]
\vspace*{-0.5cm}
                 \insertplot{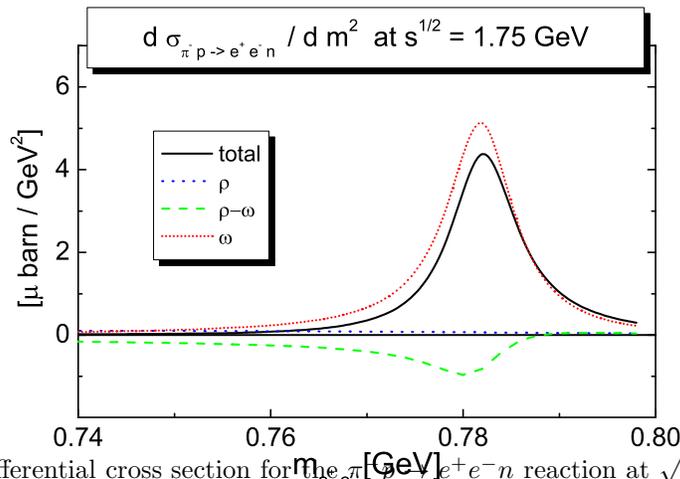}
\vspace*{-1.8cm}
\caption[]{Differential cross section for the $\pi^-p \rightarrow e^+e^- n$
reaction at $\sqrt s$=1.75 GeV as function of the invariant mass of the $e^+e^-$
pair.}
\label{fig6}
\end{figure}

\begin{figure}[h]
\vspace*{-0.5 cm}
                 \insertplot{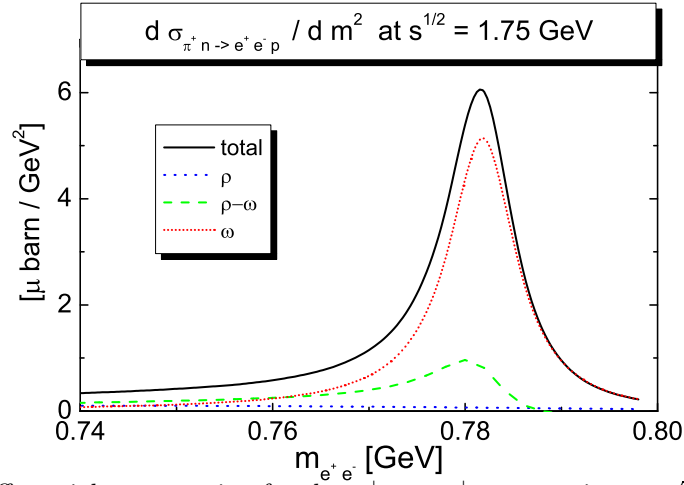}
\vspace*{-1.4 cm}
\caption[]{Differential cross section for the $\pi^+n\rightarrow e^+e^- p$
reaction at $\sqrt s$=1.75 GeV as function of the invariant mass of the $e^+e^-$
pair.}
\label{fig7}
\end{figure}

\section{Conclusion}

We have computed the $e^+e^-$ pair invariant mass distributions for the $\pi^-p \rightarrow e^+e^- n$
and $\pi^+n\rightarrow e^+e^- p$ reactions below and close to the vector meson production threshold.
We took as input the $\pi N\rightarrow\rho^0 N$ and $\pi N\rightarrow\omega N$ amplitudes
obtained in a recent relativistic and unitary coupled-channel approach to
meson-nucleon scattering \cite{Lutz1}.
Using the Vector Meson Dominance assumption, we have shown that the differential cross sections for
the $\pi^-p \rightarrow e^+e^- n$
and $\pi^+n\rightarrow e^+e^- p$ reactions below the $\omega$-threshold are very sensitive to the
coupling of low-lying baryon resonances to vector meson-nucleon
final states contributing to the $\rho^0$- and $\omega$-meson production amplitudes.
We find that the $\rho^0-\omega$ interference is destructive in the
$\pi^-p \rightarrow e^+e^- n$ channel and constructive in the $\pi^+n\rightarrow e^+e^- p$ channel
(see also Ref. \cite{Titov}).
We predict a very small cross section for the $\pi^-p \rightarrow e^+e^- n$ reaction
below threshold and a sizeable cross section for the $\pi^+n\rightarrow e^+e^- p$ reaction
in this energy range. Above the $\omega$-meson production threshold, both cross sections are
comparable and much larger.

The magnitude of the $\pi^-p \rightarrow e^+e^- n$
and $\pi^+n\rightarrow e^+e^- p$  differential cross sections below the $\omega$-threshold depends strongly
on the structure and dyna\-mics of baryon resonances. These reactions deserve
experimental studies. Such a programme could be carried at GSI (Darmstadt) using
the available pion beam and the HADES spectrometer \cite{Schoen}. These measurements
would provide a necessary step towards the understanding of $e^+e^-$ pair
production in pion-nucleus reactions and
in general significant constraints on the propagation of vector mesons in the
nuclear medium.

\end{document}